\documentclass[reprint,amsmath,amssymb,aps,pra]{revtex4-1}
\usepackage[latin9]{luainputenc}
\usepackage{float}
\usepackage{amsmath}
\usepackage{graphicx}
\usepackage{esint}

\makeatletter
\usepackage{dcolumn}
\usepackage{bm}

\makeatother

\begin{document}

\preprint{APS/123-QED}

\title{Quantum Brownian motion simulation of the control effect for two harmonic oscillators coupling in position and momentum with general environment}

\author{Hao Jia\thanks{Corresponding author: hjia4@stevens.edu}}
\author{Xiaotong Ding}
\affiliation{Department of Physics, Stevens Institute of Technology, Hoboken,
NJ 07030, US}

\date{\today}
\begin{abstract}In this paper, we study the dynamical properties of two coupled quantum harmonic
oscillators coupled with bosonic non-Markovian environment both in position and momentum. 
We deduce the exact analytical master equation using Quantum State Diffusion method and 
give the quantum trajectory description when the control is added to the system by 
applying interaction between two harmonic oscillators. With numerical simulation, we compare
the evolution of entanglement under different controlling effects. At last, we use nonlinear QSD method to strengthen our above results by getting the same evolution.
\end{abstract}
\maketitle

\section{Introduction}

\label{sec:int} The study of quantum open systems are increasingly
important in the field of quantum dynamics because it is impossible
to isolate the system from its environment or make a measurement without
involving with other systems. Generally, although environment and
specific quantum system are initially independent, they will become
entangled due to the interaction, as a result, the quantum system
will no longer be pure state which means the evolution operator is
non-unitary. Most experimental physicists face the quantum open system
where a small system of interest coupled to a large system with a
large number of freedom, which can be described by heat bath. Traditionally,
we describe the open system by a Lindblad master equation which can
be derived with Born-Markov approximation which means the flow of
energy or information is unidirectional, in other words, the bath
is memoryless. However, if the bath memory effects are relevant, for
example in the cases of a high-Q cavity, atom laser or complex structured
environment, where the Born-Markov approximation does not work, we
have to use Non-Markovian process to describe the quantum system. 

Quantum Brownian Motion (QBM) is a paradigm of quantum open system
motivated by possible observation of macroscopic effects in quantum
systems and problem of quantum measurement theory. Also, Quantum Brownian
Motion, as a exactly-solvable model, provides us a glance at the relationship
between different measure of quantum systems, including entanglement,
coherence, purity, during the evolution of quantum system and under
the external time-dependent control. To study the quantum-to-classical
transition in quantum cosmology, Hu, Paz and Zhang got the exact master
equation with nonlocal dissipation and colored noise in a general
environment, which beginning the new stage to treat the old problem\cite{HPZ}.
Iater, Chou, Ting and Hu derived an exact master equation for two
coupled quantum harmonic oscillators interacting via bilinear coupling
with a common environment at arbitrary temperature made up of many
harmonic oscillators with a general spectral density function\cite{yu2ho},
which makes it possible to study the decoherence and disentangle
in Brownian motion model. Traditionally, we use reduced density matrix
to describe the quantum open system when we consider the environment
effect to the specific system. Recently, there are tremendous progresses
in the development of stochastic Schrodinger equations to describe
the quantum open system. We will have the reduced density matrix by
tracing over the quantum trajectories of a stochastic Schrodinger
equation, which means a possible series of influence of the environment
to the system. Quantum state diffusion method provides not only an
efficient way in the numerical calculation of quantum open system,
but also a way to describe our system, which shed light on the difficulties
encountered in environment memory effect. 

In this paper, we mainly give the numerical simulation and derivation of the stochastic Schrodinger equation
and master equation for open quantum system containing two time-dependent
interacting harmonic oscillators, coupled with a thermal bath involving
infinite number of bosonic oscillators at zero temperature. The symmetric position-momentum coupling pattern is used, which also can be
regarded as a Rotating Wave Approximation of position.  We also shows in
zero temperature case, the symmetric coupling in position and momentum
provides an easy but effective way to have an glance at the quantum
system under the influence of environment and external control field,
especially for numerical simulations, because ten related differential
equations are simplified to one differential equation.

Our paper is organized as follows. We firstly give a brief introduction
to Quantum State Diffusion method, in zero temperature in Section \ref{sec:intr}.
The evolution of two harmonic
oscillators with the symmetric coupled pattern in position and momentum
in Section \ref{sec:rwa} is considered, where the time-local, convolution-less
master equation, derived by non-Markovian quantum state diffusion
method(NMQSD). We then consider the application of quantum control
in Section \ref{subsec:qc} where we control the entanglement and coherence
of the specific quantum system by time-dependent interaction. In Section \ref{subsec:num}, we simulate these controlling and non-Markov process,
 including coherence states, folk states and cat states under the influence of environment and control field.

\section{Theoretical Framework}

\subsection{Introduction to QSD in zero temperature}

\label{sec:intr} The standard total Hamiltonian in the system-plus-reservoir
in quantum open system can always be written as 

\begin{align}
\begin{aligned}H_{tot}=H_{sys}+H_{int}+H_{bath}\\
=H+\hbar\sum_{\lambda}\left(g_{\lambda}^{*}Lb_{\lambda}^{\dagger}+\text{\textit{\ensuremath{g_{\lambda}}}}L^{\dagger}b_{\lambda}\right)+\sum_{\lambda}\hbar\omega_{\lambda}b_{\lambda}^{\dagger}b_{\lambda}
\end{aligned}
\end{align}

where $L$ is the system operator providing the coupling between the
system and the environment and $g_{\lambda}$ are coupling constants.
we will then set $\text{\ensuremath{\hbar}}=1$.

Using the Schrodinger equation, the non-Markov Quantum State Diffusion
(NMQSD) equation is \cite{yu1ho}

\begin{align}
\ensuremath{\partial_{t}\psi_{t}=-iH\psi_{t}+Lz_{t}^{*}\psi_{t}-L^{\dagger}\int_{0}^{t}\alpha(t-s)\text{ }\frac{\delta\psi_{t}}{\delta z_{s}^{*}}ds}
\end{align}
where $\alpha(t-s)$ is the bath correlation function determinant
by temperature and initial bath states, $H$ is the system Hamiltonian
and $z_{t}^{*}$ are Gaussian random process with correlation functions
that mirror the vacuum correlations of the bath operators in the interaction
picture.

It is possible to replace the functional derivative by some time-dependent
operator $O$, which depends on the time $t$, $s$, and the entire
history of the stochastic process $z_{t}^{*}$. by making an ansatz,
the evolution equation for operator $O=O\left(t,s,z^{*}\right)$ is 

\begin{align}
\begin{aligned}\partial_{t}O=\left[-iH+Lz_{t}^{*}-L^{\dagger}\bar{O}\left(t,z^{*}\right),O\right]\\
-L^{\dagger}\frac{\delta\bar{O}\left(t,z^{*}\right)}{\delta z_{s}^{*}}
\end{aligned}
\end{align}
where the time-integrated operator $\bar{O}\left(t,z^{*}\right)$
is the integral of $O\left(t,s,z^{*}\right)$ from 0 to t with the
weight function$\alpha(t-s)$
\begin{align}
\ensuremath{\bar{O}\left(t,z^{*}\right)=\int_{0}^{t}\alpha(t-s)O\left(t,s,z^{*}\right)ds}
\end{align}
$\alpha(t-s)$ is the correlation function to describe the influence
of the environment to the system, which depends on the spectral density
of the system and temperature.

\begin{align}
\begin{aligned}\alpha(t-s)=\sum_{n=1}^{N}\frac{G_{n}^{2}}{2m_{n}\omega_{n}}[coth(\frac{\omega_{n}}{2k_{B}T})cos\omega_{n}(t-s)\\
-isin\omega_{n}(t-s)]
\end{aligned}
\end{align}
Our numerical results focus on zero temperature and finite temperature.
We can introduce the spectral density of bath oscillators: 
\begin{align}
J(\omega) & =\sum_{n=1}^{N}\frac{G_{n}^{2}}{2m_{n}\omega_{n}}\delta(\omega-\omega_{n})=M\gamma\omega^{3}f_{c}(\frac{\omega}{\Lambda}).
\end{align}
Later we choose super-Ohmic environment in our numerical computation.
$f_{c}$ is a cutoff function, of which $\Lambda$ is cutoff frequency,
the most important parameters to control our environmental spectrum.
In our simulation, cutoff function is chosen to be $f_{c}(x)=exp(-x)$.

The problem of solving stochastic Schrodinger equation is more and
less solving $O$ operator, exactly or approximately. The general
QSD equation can be simplified to a form which may be simulated numerically,
as long as the ansatz satisfies the consistency condition. In practice,
even when the operator $O\left(t,s,z^{*}\right)$ cannot be determined
exactly, perturbation techniques may be used to create a serious of
$O$ operator allowing for an approximate form of general QSD equation.
Fortunately, we can find exact $O$ for two coupled harmonic oscillators.
The convolution-less form of NMQSD equation will appear if we replace
the functional derivative by the known operator

\begin{align}
\ensuremath{\partial_{t}\psi_{t}=\left(-iH+Lz_{t}^{*}-L^{\dagger}\bar{O}\left(t,z^{*}\right)\right)\psi_{t}\left(z^{*}\right)}
\end{align}
Once we have the NMQSD equation, the reduced density operator is given
by the ensemble mean over the trajectories of the stochastic Schrodinger
equation. 

\begin{align}
\nonumber
\ensuremath{\dot{\rho_{t}}=-i\left[H,\rho_{t}\right]+\left[L,\mathcal{M}\left\{ P_{t}\bar{O}^{\dagger}\left(t,z^{*}\right)\right\} \right]+\\
\left[\mathcal{M}\left\{ O\left(t,z^{*}\right)P_{t}\right\} ,L\right]}\label{eq:fullms}
\end{align}
If the exact operator $O$ is independent of the noise $z_{t}^{*}$,
we can find that 

\begin{align}
\ensuremath{\dot{\rho_{t}}=-i\left[H,\rho_{t}\right]+\left[L,\rho_{t}\bar{O}^{\dagger}(t)\right]+\left[\bar{O}(t)\rho_{t},L\right]}\label{eq:rdeq}
\end{align}
In this paper, we will show that the exact master for two harmonic
oscillators in zero temperature is Eq.\ref{eq:fullms}
form while the exact master equation of Rotating Wave Approximation
form in zero temperature is Eq. \ref{eq:rdeq} form.
\subsection{Quantum Brownian motion with coupling symmetric in position and momentum
by Quantum State Diffusion}
Quantum Brownian motion of a damped harmonic oscillator
bi linearly coupled to bath of harmonic oscillators has been studied
for decades and the non-Markov exact master equation was provided
by path integral techniques\cite{yu2ho}. This part will provide
another approach to exact master equation by Quantum State Diffusion.
In the more generalized forms, the masses and frequencies of the oscillators
are different so that the master equation can be used easily in the
research of non resonant problems. Also, the couple between two harmonic
oscillators is a function of time to research the quantum control
application, and if we are not interested in the control part we can
set $k(t)=k$.

The Hamiltonian of the total system consisting of a system of two
mutually coupled harmonic oscillators with different mass and frequency
interacting with a bath of harmonic oscillators of masses and frequencies
in an equilibrium state at zero temperature:

\begin{align}
H_{tot} & =H_{sys}+H_{int}+H_{bath}
\end{align}
where

\begin{align}
\nonumber
\ensuremath{H_{sys}=\frac{p_{1}^{2}}{2M_{1}}+\frac{1}{2}M_{1}\Omega_{1}^{2}q_{1}^{2}+\frac{p_{2}^{2}}{2M_{2}}\\+\frac{1}{2}M_{2}\Omega_{2}^{2}q_{2}^{2}+k(t)\left(q_{1}-q_{2}\right){}^{2}}
\end{align}
is the system Hamiltonian for the two system oscillators of interest,
with displacements and momentum and time dependent couple function
$k(t)$.

\begin{align}
\ensuremath{H_{bath}=\sum_{n=1}^{N_{B}}\left(\frac{\pi_{n}^{2}}{2m_{n}}+\frac{1}{2}m_{n}\omega_{n}{}^{2}b_{n}^{2}\right)}
\end{align}
is the bath Hamiltonian with displacements and conjugate momentum.
\label{sec:rwa}
\\One solvable model is that
the system and the environment are coupled through different observables:

\begin{align}
\ensuremath{H_{int}=\left(q_{1}+q_{2}\right)\sum_{n=1}^{N_{B}}C_{n}b_{n}+\left(\frac{p_{1}+p_{2}}{M\Omega}\right)\sum_{n=1}^{N_{B}}\frac{C_{n}}{m_{n}\omega_{n}}\pi_{n}}
\end{align}
For simplification, we assume the masses and frequencies of those
harmonic oscillators are same. It is easy to rewrite this interaction Hamiltonian in creation and
annihilation operators\cite{paz2ho}

\begin{align}
\ensuremath{H_{int}=\sum_{n=1}^{N_{B}}\frac{C_{n}2\sqrt{2}}{\sqrt{Mm_{n}\Omega\omega_{n}}}\left(ab_{n}^{\dagger}+a^{\dagger}b_{n}\right)}\label{eq:rwa}
\end{align}
The same type can also be derived from standard position-position coupling with rotating-wave
approximation (RWA). In the expansion
of $(q-q')^{2}$, we can rewrite this in terms of creation and annihilation
operators as $(a_{1}-a_{2}+a_{1}^{\dagger}-a_{2}^{\dagger})^{2}$.
In RWA, high-frequency oscillating terms like $a^{2}$and $a^{\dagger2}$
are neglected. If we rewrite the system with RWA interaction as:\\\\

\begin{align}
\begin{aligned}H_{tot}=\Omega\left(a_{1}^{\dagger}a_{1}+a_{2}^{\dagger}a_{2}\right)+k(t)\left(a_{1}-a_{2}+a_{1}^{\dagger}-a_{2}^{\dagger}\right){}^{2}\\\sum_{n=1}^{N_{B}}\omega_{n}b_{n}^{\dagger}+\sum_{n=1}^{N_{B}}(g_{n}a_{1}b_{n}^{\dagger}+g_{n}a_{2}b_{n}^{\dagger})\\
+\sum_{n=1}^{N_{B}}(g_{n}a_{1}^{\dagger}b_{n}+g_{n}a_{2}^{\dagger}b_{n})
\end{aligned}\end{align}
In this model, the Lindblad operator will be $a_{1}+a_{2}$. We find
the $O$ operator without an explicit noise dependence:

\begin{align}
\ensuremath{O\left(t,s,z^{*}\right)=f(t,s)\left(a_{1}+a_{2}\right)}
\end{align}
Define the integration of function $f$ with the correlation function
as weight function

\begin{align}
\ensuremath{F(t)=\int_{0}^{t}f(t,s)\alpha(t-s)ds}
\end{align}
The complex function f follows the evolution equation and boundary
condition:
\begin{align}
\ensuremath{\partial_{t}f=i\Omega f+2F(t)f}
\end{align}
\begin{align}
\ensuremath{f(t,s=t)=1}
\end{align}
We can get the analytical solution if the frequency distribution is
Lorentz spectrum, where $\ensuremath{\alpha(t,s)=\frac{\Gamma\gamma}{2}e^{-\gamma|t-s|}}$.If
we integral both sides of the evolution equation of $f$ from $0$
to $t$ with $\alpha(t,s)$, we will have 
\begin{align}
\ensuremath{\partial_{t}F=2F^{2}-(\gamma-i\Omega)F+\frac{\Gamma\gamma}{2}}
\end{align}
which is Riccati Equation. 
Solve the equation with transformation $$F(t)=(-y'(t))/(2y(t))$$ and we
will have

\begin{align}
y^{\prime\prime}+\left(\gamma-i\Omega\right)y'+\Gamma\gamma y=0
\end{align}
Finally, the solution is

\begin{align}
\ensuremath{F(t)=\frac{-\lambda_{1}e^{\lambda_{1}t}+\lambda_{1}e^{\lambda_{2}t}}{2\left(e^{\lambda_{1}t}-\frac{\lambda_{1}}{\lambda_{2}}e^{\lambda_{2}t}\right)}}
\end{align}
where the parameters $\Delta=\left(\gamma-i\Omega\right)^{2}-4\Gamma\gamma$
determines whether $C(t)$ is complex or real. In this case, as long
as $\Omega\neq0$ or $\gamma^{2}-4\Gamma\gamma<0$, the system will
be non-Markov.

After obtaining the evolution of $O$, master equation turns to
be as follows: 
\begin{widetext}
\begin{align}
\partial_{t}\rho=i[\Omega(a_{1}^{\dagger}a_{1}+a_{2}^{\dagger}a_{2})+k(t)(a_{1}^{\dagger}-a_{2}^{\dagger}+a_{1}-a_{2})^{2},\rho]+[(a_{1}+a_{2}),\rho F^{*}(t)(a_{1}^{\dagger}+a_{2}^{\dagger})]+[F(t)(a_{1}+a_{2})\rho,(a_{1}^{\dagger}+a_{2}^{\dagger})].
\end{align}
\end{widetext}
\begin{figure}[H]
\noindent\begin{minipage}[t]{1\columnwidth}%
\includegraphics[width=8.5cm]{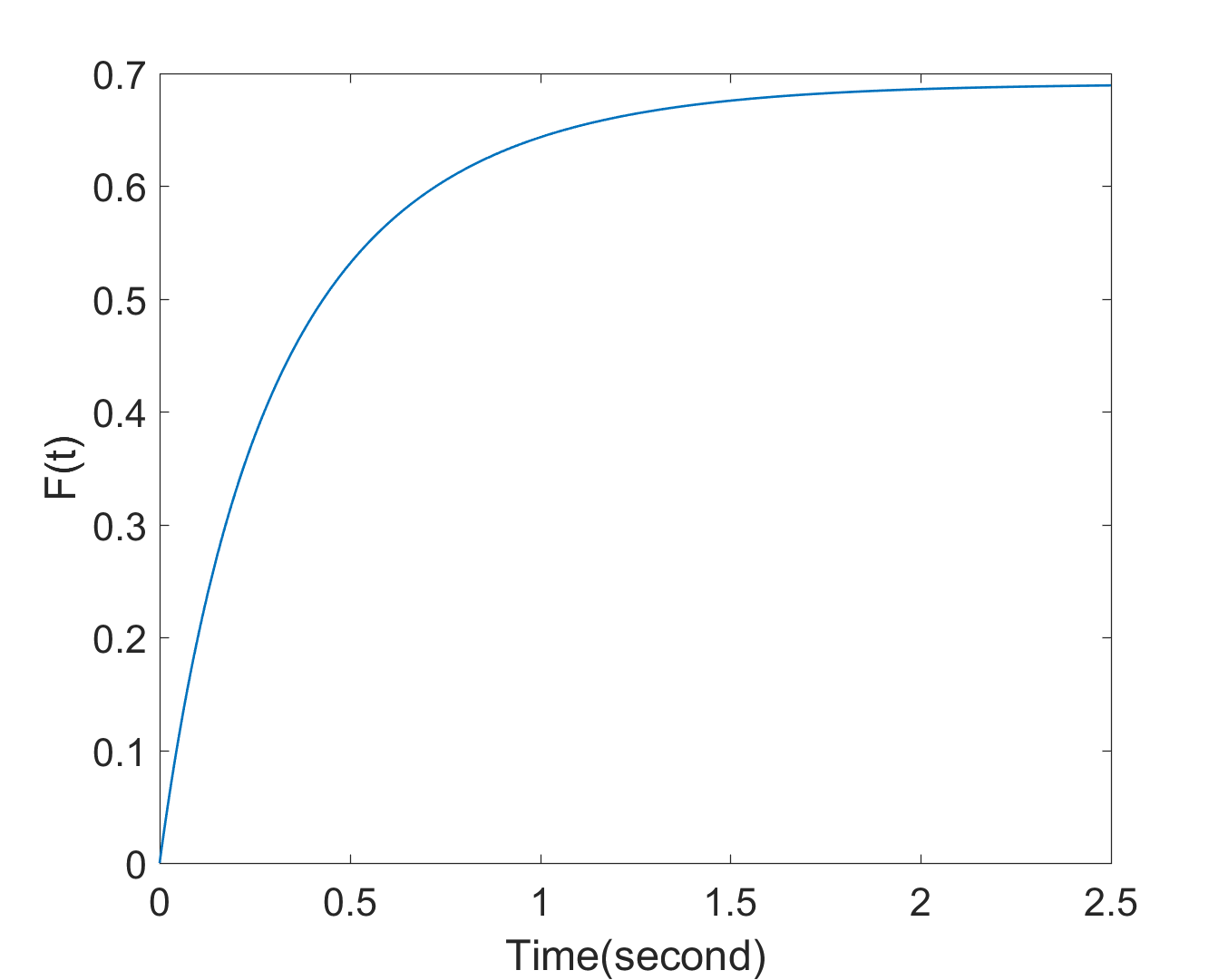} 
\end{minipage}%
\caption{If $\Delta>0(\Gamma=1,\gamma=5,\Omega=0)$, the system will become\\\hspace{\textwidth} Markov when time goes by.}
\noindent\begin{minipage}[t]{1\columnwidth}%
\includegraphics[width=8.5cm]{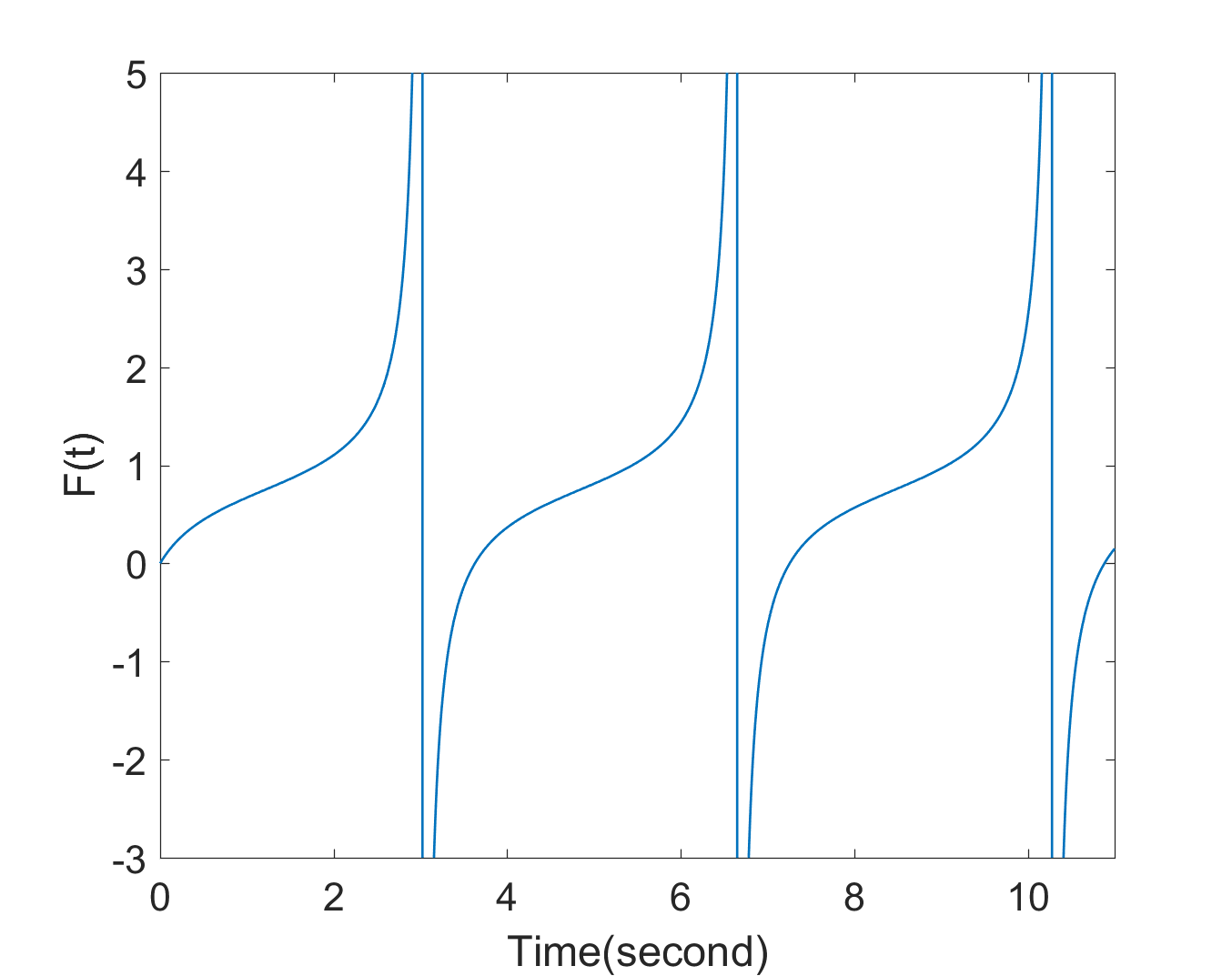}
\end{minipage}%
\caption{If $\Delta<0(\Gamma=1,\gamma=3,\Omega=0)$, system has obvious \\\hspace{\textwidth}non-Markov
properties.}
\end{figure} $\Omega,\hbar$ and $M$ have been chosen to be 1 for simplicity. It is remarkable that this case is only for zero temperature where the Landblad operator is not Hermitian, $a_{1}+a_{2}\neq(a_{1}+a_{2})^{\dagger}$. It is a good example to solve numerically because eight related integral-diffential equations have been reduced to only one integral-diffential equation $F(t)$. 
\\The resulting nonlinear, convolution less non-Markov stochastic Schrodinger
equation for the Brownian motion of a harmonic oscillator with a coupling
to the environment through position thus reads

\begin{align}
\begin{aligned}\ensuremath{\ensuremath{\frac{d|\widetilde{\psi}\rangle}{dt}=[-iH_{s}+\triangle_{t}(L)\widetilde{z}_{t}^{*}]|\widetilde{\psi}(t)\rangle-}}\\
\triangle_{t}(L^{\dagger})\bar{O}(t,\widetilde{z^{*}})|\widetilde{\psi}(t)\rangle+\triangle_{t}(L^{\dagger})\bar{O}(t,\widetilde{z^{*}})\rangle_{t}|\widetilde{\psi}(t)\rangle
\end{aligned}
\end{align}
 Here, we have the notation 
\begin{align}
\triangle_{t}(L) & \equiv L-\left\langle L\right\rangle _{t}
\end{align}
 and the noise term 
\begin{align}
\widetilde{z}_{t}^{*} & =z_{t}^{*}+\intop_{0}^{t}\alpha^{*}(t,s)\left\langle L^{\dagger}\right\rangle _{s}ds.
\end{align}

\section{Application in Quantum Control}

\label{sec:qc} In this section, we present theoretical analysis and
numerical results of the open quantum system.

\subsection{Quantum control for entanglement for gaussian states}

\label{subsec:qc}The decoherence of two harmonic oscillators has been studied systematically
in \cite{yu2ho,paz2ho} where coupling constant between two oscillators
is constant. If we control the coupling relation as a function of
time, we can control the decoherence. With the control
function, we can transform the dynamical phase from one to a different
one. or simplification, we assume the masses and frequencies of those
harmonic oscillators are same (resonant case).

It is convenient to use coordinates $x_{\pm}=\left(x_{1}\pm x_{2}\right)/\sqrt{2}$since
$x_{+}$ couples to the environment while $x_{-}$ is controlled:

\begin{align}
\begin{aligned}H_{tot}=\frac{p_{+}^{2}}{2m}+\frac{m}{2}\Omega^{2}x_{+}^{2}+\sqrt{2}x_{+}\sum_{n=1}^{N_{B}}c_{n}q_{n}\\
+\sum_{n=1}^{N_{B}}\left(\frac{p_{n}{}^{2}}{2m_{n}}+\frac{1}{2}m_{n}\omega_{n}{}^{2}q_{n}^{2}\right)\\
+\frac{p_{-}^{2}}{2m}+\frac{m}{2}\Omega^{2}x_{-}^{2}+2k(t)x_{-}^{2}
\end{aligned}
\end{align}
The system is simplified to a harmonic oscillator coupled to environment
and another harmonic oscillator controlled by external time dependent
frequency $\ensuremath{\Omega'(t)=(\Omega^{2}+4k(t)/m)^{\frac{1}{2}}}$.
The master equation for one harmonic oscillator\cite{HPZ}
is

\begin{align}
\begin{aligned}\ensuremath{\partial_{t}\rho=-i\left[H_{R},\rho\right]-i\gamma(t)\left[x_{+},\left\{ p_{+},\rho\right\} \right]}\\
-D(t)\left[x_{+},\left[x_{+},\rho\right]\right]-f(t)\left[x_{+},\left[p_{+},\rho\right]\right]
\end{aligned}
\end{align}
where the renormalized Hamiltonian is 

\begin{align}
\ensuremath{H_{R}=\frac{p_{+}^{2}}{2m}+\frac{M}{2}\Omega^{2}x_{+}^{2}+\frac{p_{-}^{2}}{2m}+\frac{M}{2}\Omega^{2}x_{-}^{2}+\frac{M}{2}\delta\omega^{2}(t)x_{+}^{2}}
\end{align}
where the coefficients depend on the spectral density of the environment.
The second moments of $x_{+}$and $p_{+}$ according to \cite{paz2ho}
are

\begin{align}
\ensuremath{\partial_{t}\left(\frac{\left\langle p_{+}^{2}\right\rangle }{2m}\right)+\frac{m}{2}\Omega^{2}(t)\partial_{t}\left\langle x_{+}^{2}\right\rangle =-2\frac{\gamma(t)}{m}\left\langle p_{+}^{2}\right\rangle +\frac{D(t)}{m}}
\end{align}

\begin{align}
\ensuremath{\frac{1}{2}\partial_{t}^{2}\left\langle x_{+}^{2}\right\rangle +\gamma(t)\partial_{t}^{2}\left\langle x_{+}^{2}\right\rangle +\Omega^{2}(t)\left\langle x_{+}^{2}\right\rangle =\frac{\left\langle p_{+}^{2}\right\rangle }{m}-\frac{f(t)}{m}}
\end{align}
Then, the system will become two separable harmonic oscillators: one
is coupled with environment and another one is under the control.
Similarly, we can have the second moments for $x_{-}$and $p_{-}$.

\begin{align}
\ensuremath{\partial_{t}\left(\frac{\left\langle p_{-}^{2}\right\rangle }{2m}\right)+\frac{m}{2}(\Omega^{2}+\frac{4k(t)}{m})^{\frac{1}{2}}\partial\left\langle x_{-}^{2}\right\rangle =0}
\end{align}

\begin{align}
\ensuremath{\frac{1}{2}\partial_{t}^{2}\left\langle x_{-}^{2}\right\rangle +(\Omega^{2}+\frac{4k(t)}{m})\left\langle x_{-}^{2}\right\rangle =\frac{\left\langle p_{-}^{2}\right\rangle }{m}}
\end{align}
We assume that the initial state of system is Gaussian. Because the
complete evolution is linear and all the operators are quadratic,
the Gaussian nature of the state will be preserved for all times.
The entanglement for Gaussian states is entirely determined by the
properties of the con-variance matrix. The covariance matrix (CM) $\sigma$
is a real symmetric and positive $4\times4$ matrix. 

\begin{equation}
\sigma(t)=\left[\begin{array}{cccc}
\sigma_{x_{1}x_{1}}(t) & \sigma_{x_{1}p_{1}}(t) & \sigma_{x_{1}x_{2}}(t) & \sigma_{x_{1}p_{2}}(t)\\
\sigma_{x_{1}p_{1}}(t) & \sigma_{p_{1}p_{1}}(t) & \sigma_{x_{2}p1}(t) & \sigma_{p_{1}p_{2}}(t)\\
\sigma_{x_{1}x_{2}}(t) & \sigma_{x_{2}p_{1}}(t) & \sigma_{x_{2}x_{2}}(t) & \sigma_{x_{2}p_{2}}(t)\\
\sigma_{x_{1}p_{2}}(t) & \sigma_{p_{1}p_{2}}(t) & \sigma_{x_{2}p_{2}}(t) & \sigma_{p_{1}p_{2}}(t)
\end{array}\right]
\end{equation}
where the matrix elements are defined as:

\begin{align}
\ensuremath{\sigma_{ij}=\frac{1}{2}\text{Tr}[(\xi_{i}\xi_{j}+\xi_{j}\xi_{i})\rho]-}\text{Tr}[\xi_{i}\rho]\text{Tr}[\xi_{j}\rho]
\end{align}
where $\xi=(x_{1},p_{1},x_{2},p_{2})$. CM also has the structure

\begin{equation}
\sigma(t)=\left[\begin{array}{cc}
A & C\\
C^{T} & B
\end{array}\right]
\end{equation}
where $A,B$ and $C$ are $2\times2$ Hermitian matrices and $T$
denotes the transposed matrix. $A$ and $B$ denote the symmetric
covariance matrices for the individual reduced one-mode states, while
the matrix $C$ contains the cross-correlations between modes. The
advantage of gaussian state is that a good measure of entanglement
for such states is logarithmic negativity $E_{N}$, which can be computed
as 

\begin{equation}
E_{N}=max\{0,-\text{ln}(2\nu_{min})\}
\end{equation}
where $\nu_{min}$ is the smallest symplectic eigenvalue of the partially
transposed covariance matrix.Some expressions for $E_{N}$ based on
previous paper \cite{paz2ho} have been known. For two mode squeezed
state, obtained from the vacuum by acting with the creation operator
$exp(-r(a_{1}^{\dagger}a_{2}^{\dagger}-a_{1}a_{2}))$, we would have
$E_{N}=2\left|r\right|$. For this squeezed state, the dispersions
satisfy the minimum uncertainty condition $\delta x_{+}\delta p_{+}=\delta x_{-}\delta p_{-}=1/2$.
The squeezing factor determines the ratio between variances since
$m\Omega\delta x_{+}/\delta p_{+}=\delta p_{-}/(m\Omega\delta x_{-})=exp(2r)$.
If $r\rightarrow\infty$, the state becomes localized in the $p_{+}$and
$x_{-}$variables approaching an ideal Einstein-Podolsky-Rosen state.
We will use numerical simulation to see how time-dependent control
external field will influence the evolution of entanglement.
\subsection{Numerical results and discussion}
\label{subsec:num}Within the framework of evolution equation, entanglement, system energy
and quantum coherence, measured by $l1$ norm can be investigated
numerically. Initially, we consider the simplest case, the symmetric
coupling between momentum and position in zero temperature, to show
how input control can result in the change of entanglement, energy
and coherence. We keep the parameters of environment, which is strong
non-Markov case, and the coupling between two harmonic oscillators
weak relative to coupling between system and environment. We simulate
system by evolute the master equation. The initial states for all
simulations are squeezed states created by squeezed operator $exp(-r(a_{1}^{\dagger}a_{2}^{\dagger}-a_{1}a_{2}))$,
we would have $E_{N}=2\left|r\right|$. Without control, the entanglement,
coherence and energy will decrease from the initial value with fluctuations,
where the non-Markov environment results in fluctuation by information
and energy back flow. In long time limit, the entanglement will stay
in a stable value. 
\begin{figure}[H]
\noindent\begin{minipage}[t]{1\columnwidth}%
\includegraphics[width=8.5cm]{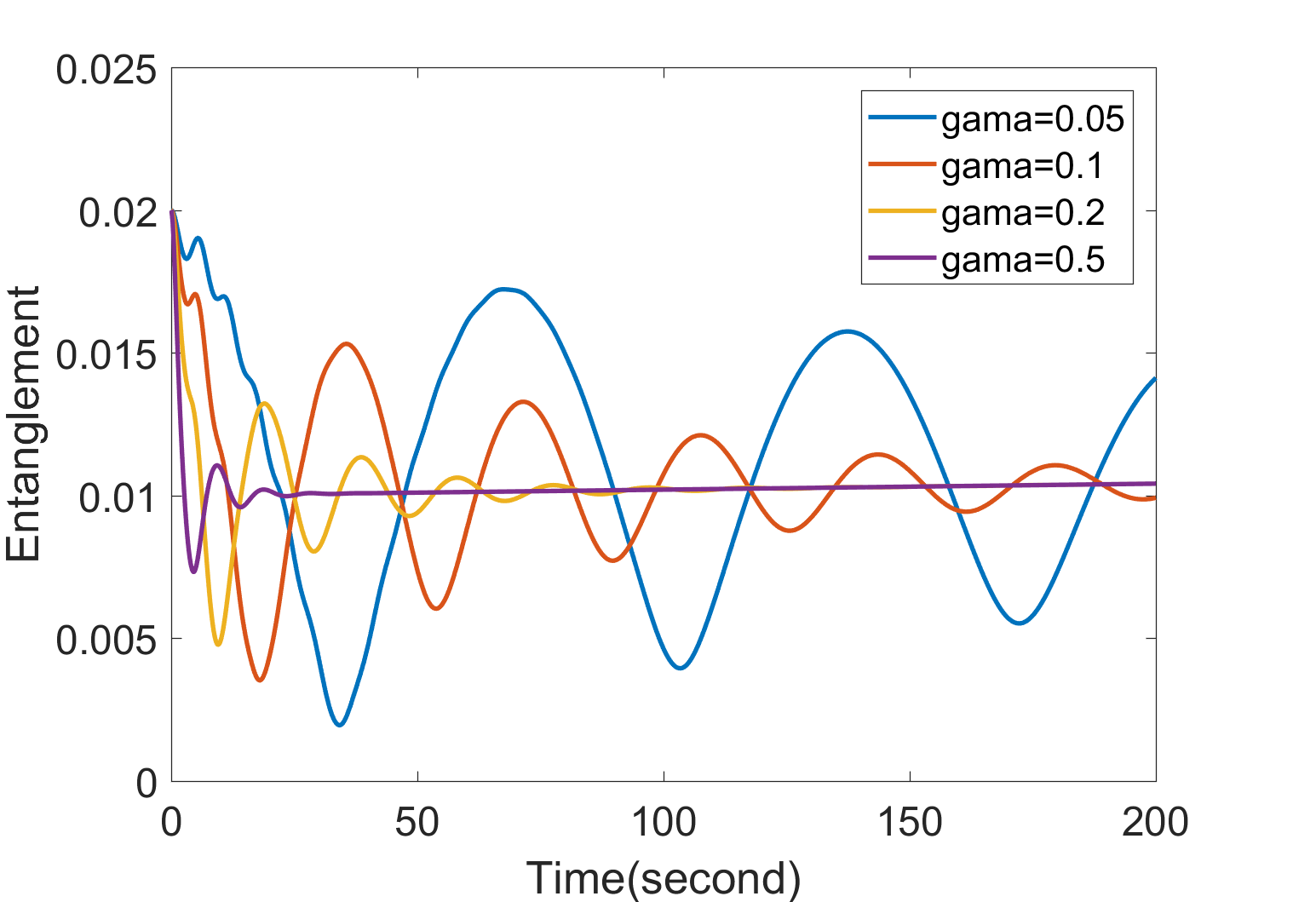}%
\end{minipage}
\caption{Entanglement of non-Markov process without control}
\end{figure}
With the control which frequency is much greater than the response
frequency, the entanglement will decrease with oscillations with high
input frequency but get stable value more quickly, which means high-frequency
input cannot influence the entanglement to reach a new point which
does not exist before the control. There are several resonance frequencies
as shown in the figure. Keeping the strength of the input signal but
changing the frequency, the oscillation of the entanglement, energy
and coherence is growing while the frequency is approaching to the
resonance frequency, and decaying while the frequency is leaving the
resonance frequency. When the input signal frequency is exactly same
as the resonance frequency, the energy will accumulate to a high level
compared with non-resonance cases and fluctuate, as what happens in
classical regime, and the analysis of classical system with comparison
between classical system and quantum system will be given later. The
coherence will increase with similar trend as energy which can be
regarded as the input of information with energy, then the flowing
of information and energy to environment is weakened by changing to
a specific resonance frequency compared to the incoming flowing information
and energy, which results in the accumulation of energy and information.There
is an interesting phenomenon for entanglement in the resonance cases,
after reaching the highest point, the entanglement drops quickly with
fluctuation and reaches zero, which means two harmonic oscillators
disentangle just after the maximum of entanglement. The entanglement
between two harmonic oscillators cannot reach zero because the environment
is non-markov except by resonance control, which is different from
classical cases. In experiment, the input signal with specific frequency
can protect the system coherence and entanglement from real environment.

\begin{figure}
\centering\includegraphics[width=8cm]{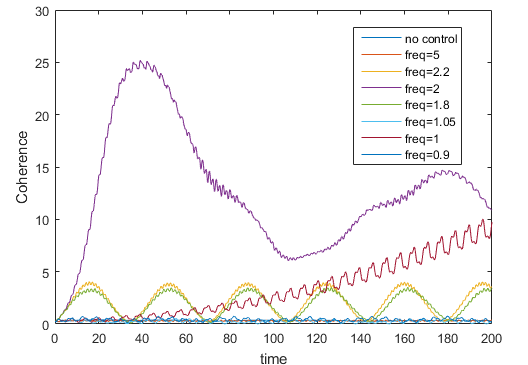}
\centering\includegraphics[width=8cm]{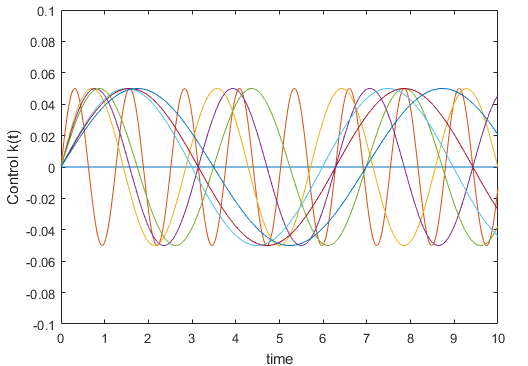}
\centering\includegraphics[width=8cm]{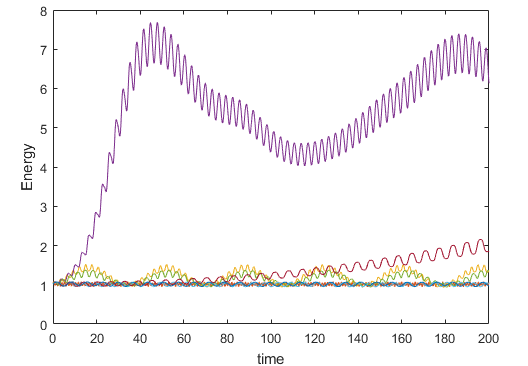}
\centering\includegraphics[width=8cm]{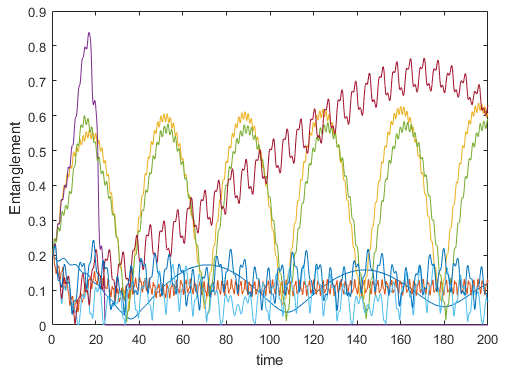}
\centering\caption{How entanglement evolutes with different environment and under different
control}
\end{figure}
In order to make our results more credible, we calculate two above-mentioned
models using two different methods, one by quantum state diffusion
equation and the other by master equation. If our above results are
valid, these two methods should give the same evolutionary picture
of entanglement over time. Finally, our comparation clearly shows
when the number of trajectories run by equation of quantum state diffusion
is large enough, it gives the highly likely result with master equation,
which shows our above numerical discussion is believable enough.
\begin{figure}[H]
\noindent\begin{minipage}[t]{1\columnwidth}%
\includegraphics[width=8.5cm]{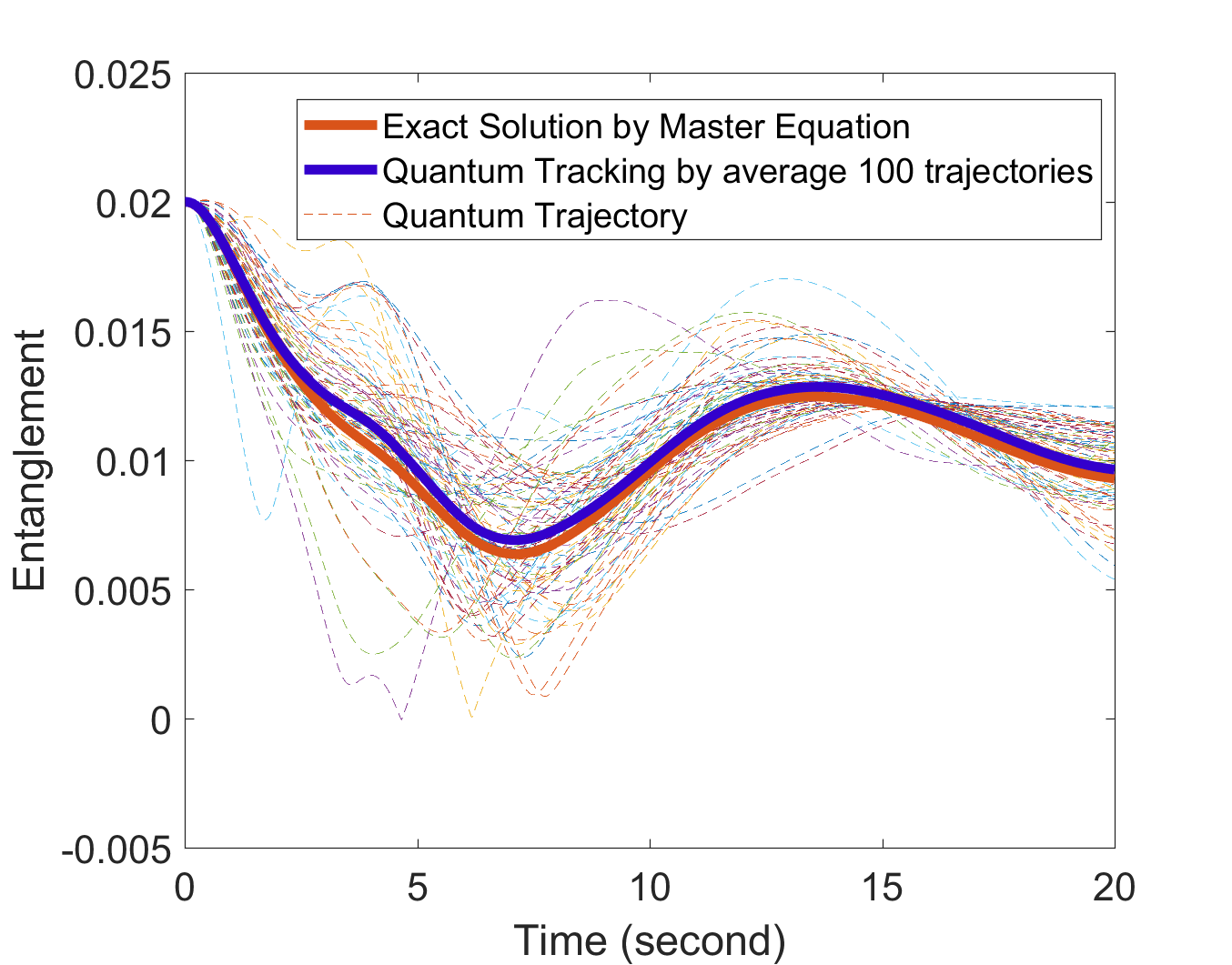}

\includegraphics[width=8.5cm]{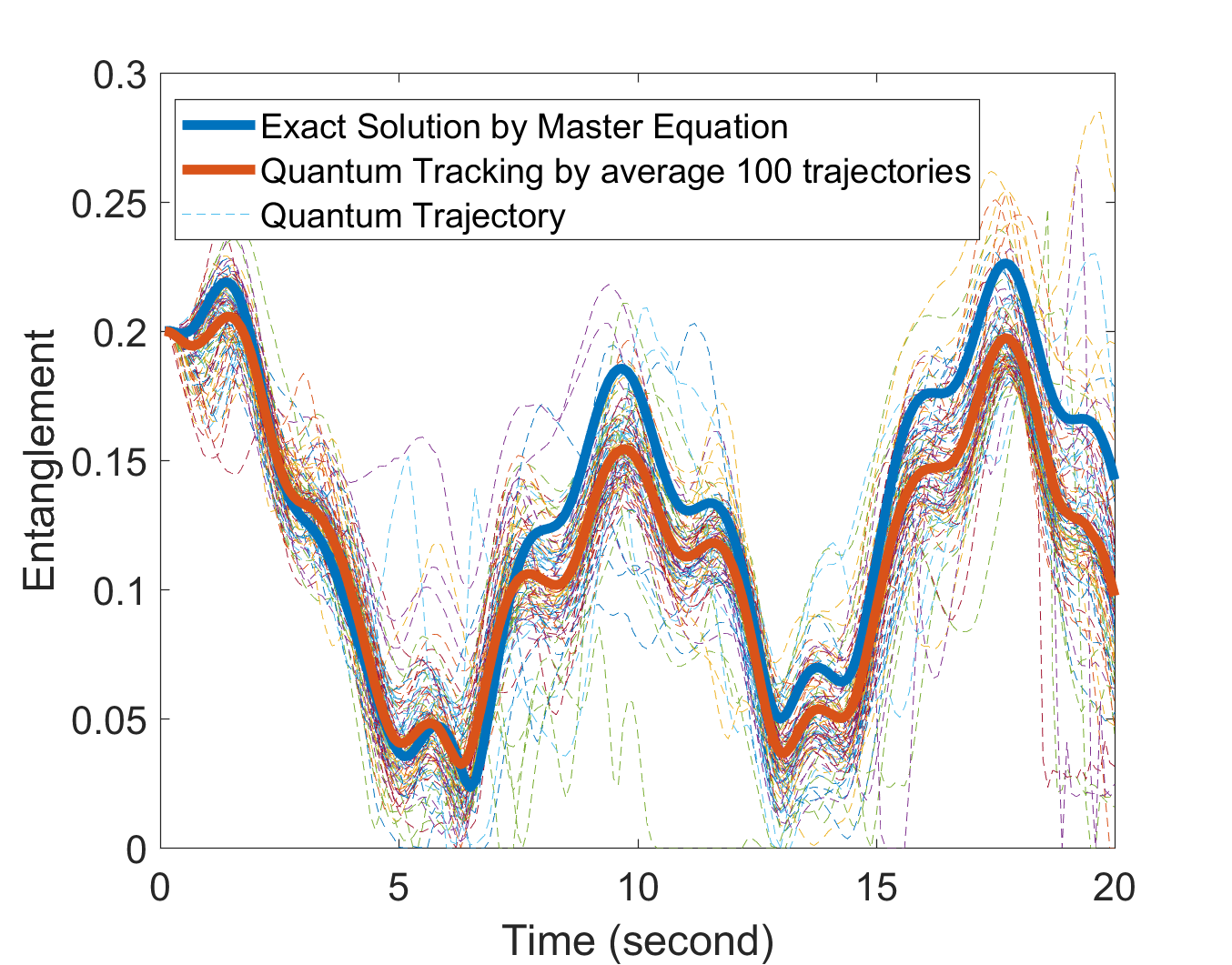}%
\end{minipage}

\caption{Comparation between Quantum State Diffusion method and Master equation
method}
\end{figure}

\section{Conclusion}

In this paper, we studied the control effect with two interacting
oscillators coupled with bosonic thermal bath by the method of quantum brownian motion. Non-Markovian master
equation of reduced density matrix is obtained by quantum state diffusion
method and one momentum-position coupling patterns are considered. There are
several attracting questions remaining in this work. We noticed that
after interacting with environment, two quantum oscillators evolve
from a pure state to a final mixed state, and correspondingly physical
quantities such as energy of system, quantum coherence and purity
remain at a non-zero level after fluctuating at the beginning. Environment
has the ability to build up coherence and entanglement but cannot
totally destroy the connection or squeeze all information contained
within system back to surroundings. Classically second law of thermodynamics
might help us to understand relative phenomenon in qualitative view
but we aim to find out the quantitative interpretation by means of
this model. We speculate that there could be some symmetry protected
process which makes the system get rid of thorough destruction in
long-time limit. Relevant investigation will be presented in future
work.
\\
\\
\\
\\\\\\\\\\
\\\\
\\\\\\\\\\

\appendix

\bibliographystyle{plain}
\bibliography{paper}
 
\end{document}